\documentclass{emulateapj}

\shorttitle{Single-lined Spectroscopic Binary Star Candidates in the RAVE Survey}
\shortauthors{Matijevi\v c et al.}

\begin{document}
\title{Single-lined Spectroscopic Binary Star Candidates in the RAVE Survey}

\author{
G.~Matijevi\v c\altaffilmark{1}, 
T.~Zwitter\altaffilmark{1,2},
O.~Bienaym\' e\altaffilmark{3},
J.~Bland-Hawthorn\altaffilmark{4}, 
K.~C.~Freeman\altaffilmark{5},  
G.~Gilmore\altaffilmark{6},    
E.~K.~Grebel\altaffilmark{7}, 
A.~Helmi\altaffilmark{8},       
U.~Munari\altaffilmark{9},
J.~F.~Navarro\altaffilmark{10},
Q.~A.~Parker\altaffilmark{4,11},   
W.~Reid\altaffilmark{11},
G.~M.~Seabroke\altaffilmark{12}, 
A.~Siebert\altaffilmark{3},
A.~Siviero\altaffilmark{9,13},
M.~Steinmetz\altaffilmark{13},
F.~G.~Watson\altaffilmark{3}, 
M.~Williams\altaffilmark{13}, and
R.~F.~G.~Wyse\altaffilmark{14}
}
\affil{$^1$University of Ljubljana, Faculty of Mathematics and Physics, Jadranska 19, 1000 Ljubljana, Slovenia}
\email{gal.matijevic@fmf.uni-lj.si}
\affil{$^2$Center of excellence SPACE-SI, Ljubljana, Slovenia}
\affil{$^{3}$Observatoire de Strasbourg, Universit\'e de Strasbourg, CNRS, 11 rue de l'universit\'e, 67000 Strasbourg, France}
\affil{$^4$Australian Astronomical Observatory, P.O. Box 296, Epping, NSW 1710, Australia}
\affil{$^5$RSAA, Australian National University, Canberra, Australia}
\affil{$^{6}$Institute of Astronomy, Cambridge, UK}
\affil{$^{7}$Astronomisches Rechen-Institut, Zentrum f\"ur Astronomie der Universit\"at Heidelberg,  Heidelberg, Germany}
\affil{$^{8}$Kapteyn Astronomical Institute, University of Groningen, Groningen, The Netherlands}
\affil{$^{9}$INAF Osservatorio Astronomico di Padova, Asiago, Italy}
\affil{$^{10}$University of Victoria, Victoria, Canada}
\affil{$^{11}$Macquarie University, Sydney, Australia}
\affil{$^{12}$Mullard Space Science Laboratory, University College London, Holmbury St Mary, Dorking, RH5 6NT, UK}
\affil{$^{13}$Leibniz-Institut für Astrophysik Potsdam (AIP), An der Sternwarte 16, 14482 Potsdam, Germany}
\affil{$^{14}$John Hopkins University, Baltimore, Maryland, USA}

\begin{abstract}

Repeated spectroscopic observations of stars in the Radial Velocity Experiment 
(RAVE) database are used to identify and examine single-lined binary (SB1) candidates.
The RAVE latest internal database (VDR3) includes 
radial velocities, atmospheric and other parameters for approximately quarter 
million of different stars with little less than 300,000 observations. In the 
sample of $\sim$20,000 stars observed more than once, 1333 stars with 
variable radial velocities were identified. Most of them are believed to be SB1 candidates.
The fraction of SB1 candidates among stars with several observations is
between 10\% and 15\% which is the lower limit for binarity among RAVE stars.
Due to the distribution of time spans between the re-observation that is biased 
towards relatively short timescales (days to weeks), the periods of the 
identified SB1 candidates are most likely in the same range.
Because of the RAVE's narrow magnitude range most of the dwarf candidates belong to the 
thin Galactic disk while the giants are part of the thick disk with distances 
extending to up to a few kpc. The comparison of the list of SB1 candidates to the VSX 
catalog of variable stars yielded several pulsating variables among the 
giant population with the radial velocity variations of up to few tens of 
$\mathrm{km\, s^{-1}}$. There are 26 matches between the catalog of spectroscopic
binary orbits ($S_{B^9}$) and the whole RAVE sample for which the given periastron time and the time of RAVE 
observation were close enough to yield a reliable comparison. RAVE measurements of radial velocities 
of known spectroscopic binaries are consistent with their published radial velocity curves.

\vspace{1cm}
\end{abstract}

\keywords{binaries: spectroscopic --- methods: data analysis --- surveys}

\section{Introduction}

Binary stars are not uncommon. Many studies searching for multiplicity among 
field and cluster stars of different spectral types report that the fraction of 
binary stars in the observed sample is as high as 50\% or more for certain
spectral types. Multiplicity among dwarfs and subdwarfs was studied by 
\citet{1991A&A...248..485D,1992ApJ...396..178F,2006ApJ...640L..63L,2010ApJS..190....1R}, massive 
binary stars by \citet{2009AJ....137.3358M,2009MNRAS.400.1479S} and binary stars 
in clusters by \citet{1999ApJ...521..682A,2009A&A...493..947S}, and others. 
Same findings are also supported by numerical simulations of star cluster 
formation \citep{2009MNRAS.392..590B}. 

In the last few decades many new spectroscopic binaries were discovered, owing a 
large part to the successful CORAVEL and CfA speedometers and other instruments. 
There were several surveys dedicated to the search of spectroscopic binaries 
\citep[among others]{2002AJ....124.1144L,2006MNRAS.371.1159G,2007A&A...473..829M}. 
The larger scale Geneva-Copenhagen Survey \citep{2004A&A...418..989N} aiming at the
F and G type dwarfs was also 
successful in identification of spectroscopic binaries with a fraction of 19\%
in the observed sample. Many of the discovered binaries from various 
catalogs along with their orbital parameters are compiled by 
\citet{2004A&A...424..727P} in the $S_{B^9}$ catalog of spectroscopic binary 
orbits.

With the development of optical-fiber spectrographs that enable the observation 
of up to several hundred stars simultaneously, efficient large-scale spectroscopic surveys 
became possible. Among the largest such sky surveys with the focus on stellar 
objects are the Sloan Digital Sky Survey (SDSS) covering the northern sky and the
Radial Velocity Experiment \citep[RAVE;][]{2006AJ....132.1645S} covering the southern sky with a higher 
spectral resolution but narrower wavelength range than the SDSS. Binary star candidates 
observed as a part of the SDSS are discussed in \citet{2005A&A...444..643P}. 
Double-lined spectroscopic binary (SB2) candidates discovered in the second data 
release of the RAVE survey where published in \citet{2010AJ....140..184M} --
Paper I hereafter. The analysis of single-lined spectroscopic binary (SB1) 
candidates identified in the RAVE's latest internal database (VDR3) is the 
subject of this paper. Section \ref{section_RV} gives an overview of radial 
velocity (RV) acquisition in the RAVE survey. Section \ref{section_rep_obs} discusses 
the sample of RAVE stars that were observed multiple times and the procedure that 
was used to identify potential SB1 objects. In Section \ref{section_cat}, a 
catalog of SB1 candidates is presented. The concluding section summarizes the main results and discusses future 
work to use the discovered binary candidates for population analysis 
and to assess their influence on the Galactic RV
distribution.

\section{RAVE Radial Velocities}\label{section_RV}

RAVE is an ongoing multi-fiber spectroscopic survey based on observations with 
the UK Schmidt Telescope at the Australian Astronomical Observatory (AAO) with the goal 
of observing up to 1 million stars  in the magnitude range between $9<I<12$. The 
wavelength range of the spectra covers the near-infrared region 
$\lambda\lambda8410-8795$ with a resolving power of $R\sim 7500$, typically with 
a moderately high signal-to-noise ratio (S/N; mean value of 45 pixel$^{-1}$). The 
RAVE spectral analysis pipeline is designed to derive accurate stellar RVs
 as well as atmospheric parameters and chemical composition 
\citep{2010AAS...21545609B}. So far, two data releases have been published 
\citep{2006AJ....132.1645S,2008AJ....136..421Z} and the third one was submitted
\citep{2011temp...S}.

The measurement of stellar RVs is one of the major goals of the RAVE 
survey. Along with the star's position on the sky, its proper motion and distance, 
an RV gives us a missing sixth kinematic component. Having full six-dimensional data 
for up to a million stars allows one to carry out detailed
studies of structure and kinematics of the Galaxy.

All data provided by the RAVE survey are acquired at the AAO 1.2m UK Schmidt 
telescope with the 6dF optical fiber system feeding the dedicated spectrograph. 
About a hundred optical fibers are positioned in the focal plane of the 
telescope to route light from the same number of stars simultaneously to 
the slit vane mounted in the spectrograph enclosure. Additional fibers 
are used to observe the background sky.
The slit vane stacks the endings of all fibers in a column, one over the other. Due to the close spacing 
of the fibers there is a chance of light cross-talk between them. Light 
from one fiber may be spilling into the adjacent one. This effect is minimized by observing stars in a narrow magnitude 
range and by data reduction pipeline which subtracts the cross-talk by an 
iterative procedure. Second order light which contaminates spectra from the first RAVE data 
release is now blocked by a blue blocking filter placed in front of the 
collimator. The light passing the collimator is dispersed using 
a volume phase holographic transmission grating and finally recorded with a camera using a CCD 
detector. Because the throughput of the different fibers may vary, the final 
$\mathrm{S/N}$ of two equally bright stars from the same field plate may 
be different. Typically, fields are observed in the following manner. Before and after five 
consecutive field exposures lasting 10 minutes each, two calibrating lamp 
exposures are usually made. Slight temperature variations in the spectrograph 
enclosure may shift the wavelength calibration so each plate has a few dedicated sky 
fibers that (together with sky velocities derived 
from stellar spectra) serve as additional  wavelength constraint and are also used for the 
subtraction of the telluric lines present in stellar spectra 
\citep{2006AJ....132.1645S}.

The measurement of RVs is performed in a following way. First, a 
spectrum is continuum normalized using a pair of cubic splines. Then, a four 
step iterative method is used to calculate its RV: 
(1) using a standard cross-correlation procedure 
\citep{1979AJ.....84.1511T} a first RV estimate is calculated. This step 
requires a library of template synthetic spectra against which a correlation is 
calculated. A subset of 40 spectra from \citet{2005A&A...442.1127M} covering the 
whole $\{T_\mathrm{eff},\log g\}$ parameter space is used to get a rough estimate with 
a precision of $\sim 5\,\mathrm{km\ s^{-1}}$ for spectra with strong \ion{Ca}{2} 
lines. (2) The spectra are shifted relative to the rest frame according to the RV 
estimate just calculated. (3) A best-matching template is constructed 
from a full library of template spectra (see \citet{2008AJ....136..421Z} for a 
more detailed description of this procedure). (4) With a best-matching 
template a new RV is calculated, and finally this measurement is corrected for a 
possible zero-point shift \citep{2011temp...S} that is produced by the temperature 
variations in the spectrograph components. Errors of the RV measurements are 
estimated by fitting a parabolic curve to the peak of the cross-correlation function.

\begin{figure}
\plotone{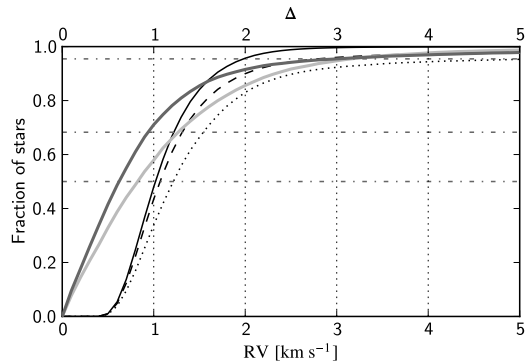}
\figcaption{Cumulative distribution of measured radial velocity errors. The 
dashed line represents the data from the VDR3 release with 
$\mathrm{S/N}>20$. The thinner black line shows a subset that includes only stars that were 
confidently classified as cold $(T_\mathrm{eff}\lesssim 7500\,\mathrm{K})$ normal 
single stars. The dotted line represents only the already published data from the 
second data release. Thicker light gray line shows the distribution of applied 
zero-point corrections (to correct the temperature variations, see the text) and the thicker 
dark gray line shows the average discrepancy between 
RVs and their weighted average in terms of RV errors (Equation~(\ref{eq_delta})) for stars with multiple observations. 
Data from the first release plagued by 
second order light contamination are excluded in all shown sets. Dash-dotted lines mark 
the 50th, 68th and 95th percentiles.
\label{plot_rv_errors}}
\end{figure}

The VDR3 database consists of 344,924 entries including the already published 
data from the first and second releases. Out of those there are 295,618 database 
entries with $\mathrm{S/N}>20$ and further exclusion of the 
second-order-light-plagued data from the first release leaves 279,120 
observations of 249,980 different stars. This subset serves as the basis of this 
paper. Repeated observations were identified by matching Two Micron All Sky Survey (2MASS) identifiers. For 
a small set of stars with missing 2MASS identifiers we compared their 
coordinates and matched stars that were not more than $5^{\prime\prime}$ apart. All matching 
cases were visually checked on the DSS plates to ensure the catalog entries 
indeed belong to the same star. We decided to exclude the data from the first 
release because the classification of those spectra is unreliable due to strong 
continuum variations and their RV measurements are usually significantly less 
accurate than the RVs of spectra recorded later. Cumulative distributions of 
RV errors for different data sets are shown in 
Figure~\ref{plot_rv_errors}. The median error of the selected subset stands at 
just over $1\, \mathrm{km\ s^{-1}}$ and RVs of 95\% of all cold single 
stars are measured to better than $2\, \mathrm{km\ s^{-1}}$. There is a notable 
improvement in RV accuracy compared to the second data release. The same figure also 
shows the distribution of applied zero-point corrections. In roughly two thirds 
of all cases this correction is smaller than the error of the measured RV,
although in some cases the correction can be an order of magnitude 
larger than the RV errors.

\section{Repeated Observations}\label{section_rep_obs}

\begin{figure}
\plotone{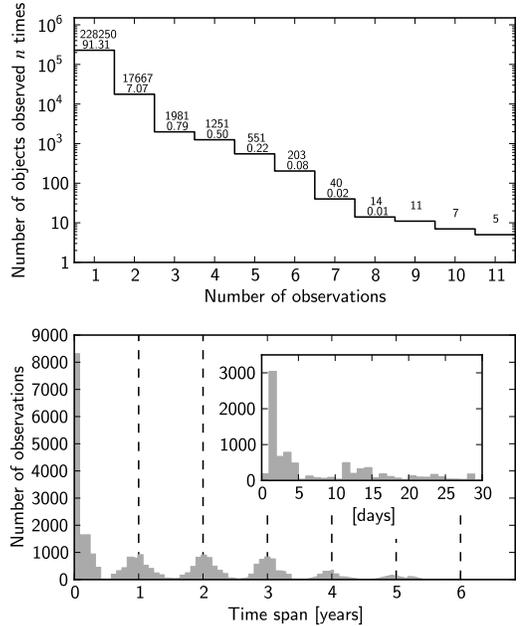}
\figcaption{Upper diagram shows the number of objects observed $n$ times 
along with the fraction that this number represents in the whole sample of 
279,120 stars. The lower diagram shows the distribution of the time span between 
the consecutive observations of the same objects. 
\label{plot_repeated_obs}}
\end{figure}

In order to identify any stars with variable RVs, more than one 
observation of the same stars is needed. The RAVE survey is primarily focused 
on the acquisition of as many different targets as possible. Re-observations of 
the same stars are done for stability and quality check purposes and also to allow
for the determination of the fraction of stars with variable RVs. 
Nevertheless, at about 9\% of all observations the number of stars that 
were observed more than once is still high (Figure~\ref{plot_repeated_obs}). 
Altogether, 21,730 stars were observed more than once in the selected subsample. 
Most of the repeats were observed in the following few days with a strong peak 
at a day after the previous observation.

The stability of RVs is shown in Figure~\ref{plot_rv_errors}. The thick dark gray 
line shows the distribution of average differences between the RV measurements 
and their weighted average ($\overline{\mathrm{RV}}$) in terms of individual errors,
\begin{equation}\label{eq_delta}
\Delta=\frac{1}{N}\sum_i^N\frac{|\mathrm{RV}_i-\overline{\mathrm{RV}}|}{\sigma_i},
\end{equation}
for all repeatedly observed normal single stars with $\mathrm{S/N}>20$ 
(same set as used for the thin solid line in Figure~\ref{plot_rv_errors}). $N$ indicates the number of observations for a given star.
A long tail of the distribution toward the larger values of $\Delta$ is indeed not an observational
error but indicates the presence of the stars with variable RVs. 

To establish a quantitative criterion for RV variability we defined the 
following function. We denoted two RVs and their errors 
measured for the same star at two different times as $(\mathrm{RV}_1,\sigma_1)$ 
and $(\mathrm{RV}_2,\sigma_2)$. The squares of the RV errors $\sigma_i^2$ can be 
treated as variances of the Gaussian distributions with the mean values 
$\mathrm{RV}_i$. If we would randomly pick two samples from each of these distributions, 
the probability that the pick from the second one is greater than 
the pick from the first one is equal to
\pagebreak
\begin{eqnarray}
P(2>1)&=&\frac{1}{\sqrt{2\pi\sigma_1^2}}\frac{1}{\sqrt{2\pi\sigma_2^2}}\\\nonumber
& & \int_{-\infty}^{\infty}\! \int_{-\infty}^{y}\!e^{-\frac{(x-\mathrm{RV}_1)^2}{2\sigma_1^2}} e^{-\frac{(y-\mathrm{RV}_2)^2}{2\sigma_2^2}}dxdy.
\end{eqnarray}
The double integral can be simplified by introducing a new variable 
$u=(x-\mathrm{RV}_1)/\sqrt{2}\sigma_1$ and evaluating the integration over $x$,
\begin{eqnarray}\label{eq_prob}
P(2>1)=&\frac{1}{\sqrt{2\pi\sigma_2^2}}&\int_{-\infty}^{\infty} \! \frac{1}{2}\left[1+\mathrm{erf}\left(\frac{y-\mathrm{RV}_1}{\sqrt{2}\sigma_1}\right)\right]\\\nonumber
& & e^{-\frac{(y-\mathrm{RV}_2)^2}{2\sigma_2^2}}dy,
\end{eqnarray}
where $\mathrm{erf}$ is the standard error function. By using the following 
identity,
\begin{equation}
\int_{-\infty}^{\infty}\! e^{-(ax+b)^2}\mathrm{erf}(cx+d)dx=\frac{\sqrt{\pi}}{a}\mathrm{erf}\left(\frac{ad-bc}{\sqrt{a^2+c^2}}\right),
\end{equation}
we can write a simple expression for the probability in question,
\begin{equation}\label{eq_erf}
P(2>1)=\frac{1}{2}\left[1+\mathrm{erf}\left(\frac{\mathrm{RV}_2-\mathrm{RV}_1}{\sqrt{2(\sigma_1^2+\sigma_2^2)}}\right)\right].
\end{equation}
If the difference in the numerator of the error function is zero (both RVs 
are the same), the probability is equal to $1/2$ and goes toward 1 for 
pairs with very different $\mathrm{RV}$s and comparably small $\sigma$s. If $\mathrm{RV}_1$ is 
greater than $\mathrm{RV}_2$, than the probability $P$ will be less than $1/2$. Since we 
are only interested if the RV changed between measurements, 
we can always label the greater of both RVs with index $2$ without the loss of generality.
The nature of error function makes it hard to tell the difference between two 
pairs with relatively large RV differences because the value of the error function 
will always be very close to $1$ in such cases. It is more convenient to define 
a new function involving the logarithm of the function $P$,
\begin{equation}\label{eq_prob_final_abs}
p_\mathrm{log}=-\log_{10}(1-P).
\end{equation}
The calculation of the logarithm in this function where the difference $1-P\approx10^{-15}$
is troublesome due to double precision floating point limitations. To prevent any numerical
inconsistencies the upper limit of the $p_\mathrm{log}$ criterion was set to 14 in the subsequent calculations. 
\begin{figure}
\plotone{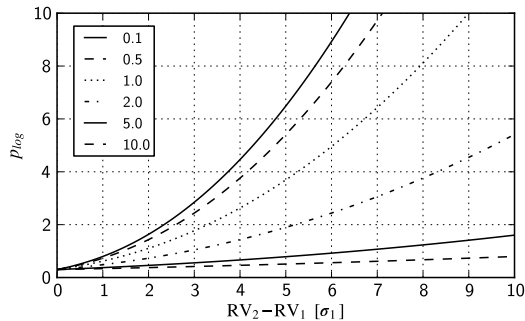}
\figcaption{$p_\mathrm{log}$ from Eq.~\ref{eq_prob_final_abs} as a function of 
difference $\mathrm{RV}_2-\mathrm{RV}_1$ in units of $\sigma_1$. Different line styles 
correspond to different values of $\sigma_2$, also in units of 
$\sigma_1$.
\label{plot_erf_prob}}
\end{figure}
The function $p_\mathrm{log}$ is shown in Fig.~\ref{plot_erf_prob} for several different error 
ratios. For example, if both errors are the same $(\sigma_1=\sigma_2)$ and the 
$\mathrm{RV}$s are roughly $4\sigma_1$ apart, the value of the function $p_\mathrm{log}$ will 
be a little less than three, corresponding to the value of $P=0.998$ so the large 
majority of the values from the second distribution are greater than the values 
from the first. The probability that the radial velocity changed from one observation 
to another is relatively high. On the other hand, if $p_\mathrm{log}<2$, the 
variability is questionable and if $p_\mathrm{log}<1$, the radial velocity variability 
is insignificant. In comparison, the lower limit of the variability criterion given 
in \citet{2005A&A...444..643P} for stars observed twice and with equal RV errors 
for both measurements is $p_\mathrm{log}=2.87$.

\section{Catalog of SB1 candidates}\label{section_cat}

We have examined a sample of 21,730 objects that were observed at least twice. 
We excluded all objects whose spectra are morphologically different from spectra 
of single stars since the described RV extraction method does not 
give reliable results in such cases, including double-lined spectroscopic 
binaries. All such objects were identified using the classification method 
described in Paper I. The method systematically examines the properties of the
cross-correlation function between each observed spectrum and a pre-calculated 
synthetic template and according to those properties groups similar spectra 
together. Among the excluded classes were spectra with observational or 
reduction errors (problematic continuum normalization, for example), hot stars 
whose Paschen series lines are too wide for precise RV measurement and 
intrinsically peculiar stars (stars with active chromospheres, emission line 
stars, etc.). The selection reduced the number of stars in the sample to 20,027.

\begin{figure}
\plotone{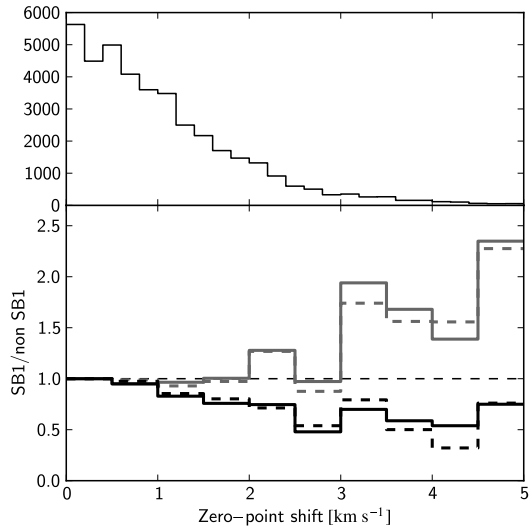}
\figcaption{Bottom diagram shows the ratio between the number of SB1 candidates vs. the number of the rest of 
repeatedly observed stars as a function of zero-point shift. All histograms are normalized
so that the first bin equals 1. Gray lines show the initial
calculation where the zero-point uncertainty is not included ($k=0$ in Equations~\ref{eq_sigma_rv_zp},\ref{eq_sigma_zp}),
whereas black lines show the same distribution for $k=0.5$. SB1 candidates in both cases
are considered all pairs with $p_\mathrm{log}>3$ (full lines) and $p_\mathrm{log}>4$ (dashed-lines).
The upper diagram shows the distribution of zero-point shifts for the pairs of observations
with the highest variability criterion for all stars observed more than once.
\label{plot_zero_point_od}}
\end{figure}

We calculated the $p_\mathrm{log}$ variability criterion for all stars from this 
sample using the RV measurements and their errors as given in the RAVE
database. In cases where more than two observations were available, the pair with 
the highest value of the criterion was considered. The inspection of a few stars with the 
largest number of repeated observations revealed that in some cases only one or two of 
the RV measurements were not consistent with the averaged RV. 
We note that some of the observations responsible 
for the RV variability have large values of the zero-point correction
which is usually very small (Section \ref{section_RV}). The comparison of the number of SB1 candidates
with zero-point shift in some interval to the rest of the repeatedly observed
stars in the same interval showed that there was a significant overdensity
of SB1 candidates in the region where at least one of the two RV measurements in the pair was
corrected for a relatively high amount (Figure~\ref{plot_zero_point_od}).
Such spurious detections would contaminate the statistics of discovered SB1 candidates.
To account for this systematic error we recalculated the variability criterion
with the adjusted error estimate:
\begin{equation}\label{eq_sigma_rv_zp}
\sigma=\sqrt{\sigma_\mathrm{RV}^2+\sigma_\mathrm{ZP}^2}, 
\end{equation}
where
\begin{equation}\label{eq_sigma_zp}
\sigma_\mathrm{ZP}=k*\mathrm{RV}_\mathrm{ZP},
\end{equation}
so the error in zero-point shift is proportional to the
shift in this approximation. Equation~\ref{eq_sigma_rv_zp} is only valid if RV and
zero-point error estimates are uncorrelated. This was checked and indeed holds true.
The coefficient $k=0.5$ was selected so that it fully eliminates the SB1 overdensity at large
zero-point shifts. It was also intentionally set high enough so that the ratio of 
SB1 candidates confirmed with the adjusted error estimate compared to 
the ones before this test is a decreasing function of the value of the 
zero-point correction. One expects that because high values of zero-point
shifts are the least trustworthy and include the highest fraction of 
potential false positives.
The overall number of SB1 candidates is not much affected by the selected value of $k$,
since the majority of all RVs have relatively small zero-point corrections (in only 5\% of the cases they
exceed $3\,\mathrm{km\ s^{-1}}$ and anyway are smaller than the RV errors, see Figure~\ref{plot_rv_errors}).

\begin{figure}
\plotone{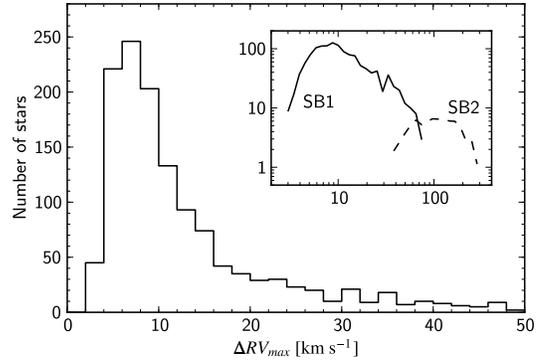}
\figcaption{Distribution of maximum changes of radial velocity between 
measurements for the identified SB1 candidates. The smaller diagram shows the 
same distribution in the logarithmic scale along with the preliminary normalized 
distribution of separations between components of identified SB2 candidates 
(dashed line).
\label{plot_rv_dist}}
\end{figure}

Altogether, 1333 or 6.6\% of the stars have $p_\mathrm{log}\geq3$ and were identified 
as potential SB1 candidates. The summary of the number of discovered candidates 
versus the number of observations is given in 
Table~\ref{table_sb1}. Since the limiting value of $p_\mathrm{log}=3$ is arbitrary,
$p_\mathrm{log}=2,4$ are given for comparison. It is evident that for $p_\mathrm{log}=2$ the number
of SB1 candidates becomes unrealistically high, making the $p_\mathrm{log}=3$ a plausible
lower limit for the variability criterion.  The efficiency of the detection is $\sim 6$\%
for twice observed star but it grows steadily toward $\sim 15$\% for stars observed five or six times. 
There are too few candidates with more 
observations to draw any conclusions. The number of identified candidates gives 
only a lower limit of spectroscopic binaries in the observed sample and 
selection criteria were indeed set restrictively to avoid as much potential 
false positives as possible and also to account for the fact that the RV error 
estimates might not always be Gaussian. With the maximal time span between 
re-observations at around 2000 days it cannot be expected that systems with 
significantly longer periods are detectable. More so, in long period systems the 
RV amplitudes become too small to be detectable and most of the SB1 candidates 
were observed for the second time only days after the first observation. The 
distribution of RV differences between measurements is shown in 
Figure~\ref{plot_rv_dist}. Most frequently the variations were small, around the value 
of $6\,\mathrm{km\ s^{-1}}$. The distribution falls quickly after 
$10\,\mathrm{km\ s^{-1}}$ but extends all the way to over $160\,\mathrm{km\ s^{-1}}$. 
All single-lined objects with high RV shifts are particularly interesting for further 
investigation due to the possibility of the presence of massive and faint objects. The 
distribution of RV shifts was also compared to the distribution of separations 
between the components of the identified SB2 candidates 
(Figure~\ref{plot_rv_dist}). The data for velocity separations were taken from the list 
of preliminary solutions for 1040 SB2 objects found in the RAVE catalog 
(partially described in Paper I). The distribution shown on the plot was scaled compared to the SB1 
distribution so that the effective number of stars from which the selection was done 
is the same for both types. The efficiency of SB2 detection becomes high at the 
point where the number of identified SB1 objects gets low.
\begin{deluxetable}{cccccccc}
\tablecaption{Number of SB1 Candidates for Different Values of $p_\mathrm{log}$\label{table_sb1}}
\tablewidth{0pt}
\tablehead{
\colhead{} & \multicolumn{2}{c}{$p_\mathrm{log}=2$} & \multicolumn{2}{c}{$p_\mathrm{log}=3$} & \multicolumn{2}{c}{$p_\mathrm{log}=4$} \\
\colhead{${N_{obs}}$} & \colhead{$N$} & \colhead{${N/N_{all}}$} & \colhead{$N$} & \colhead{${N/N_{all}}$} & \colhead{$N$} & \colhead{${N/N_{all}}$} 
}
\startdata
 2 & 1438 & 0.09 & 919 & 0.06 & 706 & 0.04\\
 3 & 263 & 0.15 & 166 & 0.09 & 127 & 0.07\\
 4 & 230 & 0.20 & 127 & 0.11 &  91 & 0.08\\
 5 & 147 & 0.28 &  71 & 0.14 &  47 & 0.09\\
 6 &  52 & 0.28 &  33 & 0.18 &  22 & 0.12\\
 7 &  17 & 0.42 &   7 & 0.17 &   5 & 0.12\\
 8 &   7 & 0.78 &   1 & 0.11 &   1 & 0.11\\
 9 &   8 & 0.73 &   4 & 0.36 &   3 & 0.27\\
10 &   6 & 0.86 &   3 & 0.43 &   2 & 0.29\\
11 &   3 & 0.75 &   2 & 0.50 &   0 & 0.00
\enddata
\end{deluxetable}

\begin{figure*}
\plotone{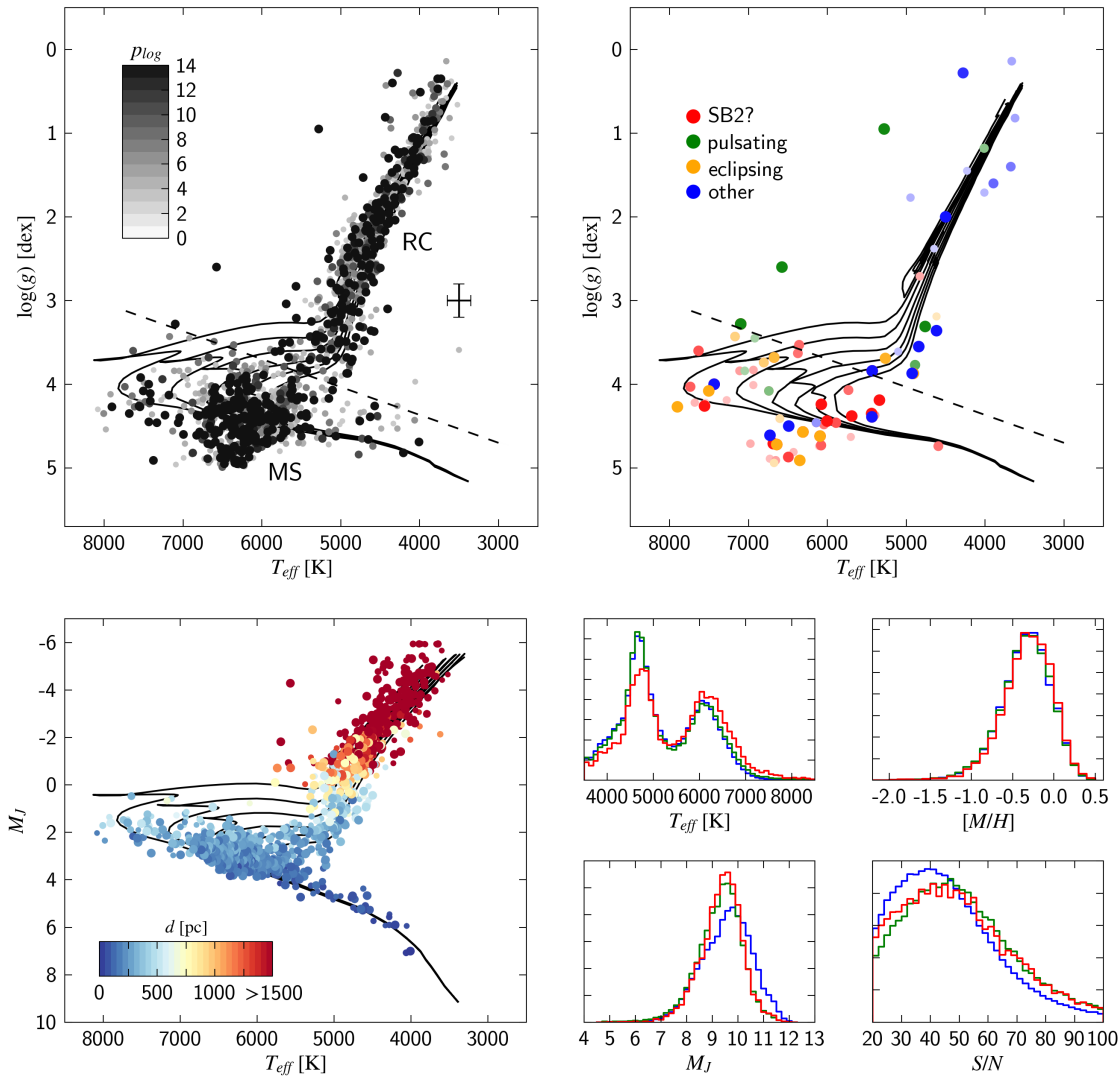}
\figcaption{{Upper left panel:} $\{T_\mathrm{eff},\log g\}$ diagram for a sample of identified SB1 
candidates. The variability criterion $(p_\mathrm{log})$ is represented with different shades of gray where darker tones 
and bigger markers correspond to greater chance of RV variability. The dashed line roughly divides 
giants mostly from the red clump (RC) region and main-sequence (MS) stars. The error bars
represent typical uncertainties of both parameters.
{Upper right panel:} photometrically variable stars from the VSX catalog 
(see the text) and suspected SB2 objects among the identified SB1 candidates. 
{Lower left panel:} SB1 candidates with color-coded distances from 
\citet{2010A&A...522A..54Z} and tone values as in the previous diagram. Here,
only the size of markers corresponds to the value of $p_\mathrm{log}$ to avoid confusion.
Isochrones for $\mathrm{[M/H]}=-0.2$ spanning $9.0\textrm{--}10.0$ in log age with 
$0.2$ steps by \citet{2008A&A...482..883M} and \citet{2010ApJ...724.1030G}. 
{Lower right panels:} normalized distributions of effective temperature, 
metallicity, 2MASS $M_J$ magnitude and S/N for non-peculiar single stars 
of the whole RAVE sample (blue), stars observed more than once (green) and SB1 
candidates (red).
\label{plot_hr}}
\end{figure*}

The $\{T_\mathrm{eff},\log g\}$ diagram of all SB1 candidates and various distributions are shown in 
Figure~\ref{plot_hr}. The distribution of effective temperature of stars observed 
multiple times has two distinct peaks at around $4500\,\mathrm{K}$ for the red 
clump stars and $6000\,\mathrm{K}$ for the main sequence dwarfs, same as in the 
overall RAVE sample. The S/N of the re-observed stars is somewhat higher 
than in the general population. The reason is that slightly brighter stars were 
observed more frequently than the faint ones. There are more SB1 
dwarf candidates than giant candidates, which is also apparent from the distribution of 
effective temperatures for the SB1 candidate sample. The ratio of giants to 
dwarfs in the whole RAVE sample is around 57:43, while the same ratio for the 
SB1 candidates is close to 50:50. This is an expected result. SB1 candidates roughly fall 
into two groups: main sequence stars with masses $\sim1-1.2\ M_\odot$ and red 
clump giants with masses slightly larger than the first group (based on isochrones
by \citet{2008A&A...482..883M}). The important difference
between the two groups is approximately an order of magnitude larger radii of giant stars.
This implies that the smallest binary orbits than can host giant stars
must be larger than the orbits of main sequence stars due to Roche lobe
radius limits (see for example \citet{1983ApJ...268..368E}). Consequently, the periods of SB1 binaries that host giant stars will be longer
and average RV shifts will be smaller and therefore harder to detect.

The distribution of metallicity is relatively wide since RAVE observes field 
stars representing the general population of thin and thick disks. The metallicity of 
SB1 candidates does not differ from the rest of the stars meaning that SB1s are 
also scattered throughout both disks. Due to RAVE's narrow magnitude range, dwarf 
candidates are mostly limited to the thin disk, while giants also reach into
the Galaxy's thick disk (lower left panel of Figure~\ref{plot_hr};
see also \citet{2010A&A...522A..54Z} for further discussion on distances).

Visual inspection of spectra revealed that 43 SB1 candidates could in fact be 
double-lined spectroscopic binaries if observed at higher resolutions or at more
favorable phases (red markers in the upper right panel of Figure~\ref{plot_hr}).
The assumption is based on their slightly asymmetric and widened lines but
the effect is too small for more confident confirmation.  
Note that atmospheric parameters for such stars are generally incorrect since the 
pipeline always assumes that the observed spectrum belongs to a single star 
and models the binary spectrum with this assumption. Nevertheless, 
the general region (dwarf/giant) of the solution is correct since the most 
affected parameter is rotational velocity while the effective temperature and 
surface gravity are not very sensitive to the apparent slight line broadening 
that is produced by the shift. Most of these stars are close to the main sequence.

We do not wish to attempt to infer binary orbits for those stars with the highest number of 
observations. While it is possible to derive orbital parameters for 
stars with six or more observations in theory, the precision of the calculated solution 
also depends on how well the measured points are distributed along the 
orbital phase, on the uncertainty of RVs and so on. We tried to fit the 
orbits for some stars, but none of the solutions turned out to 
be reliable.

The catalog of all stars with variable RVs will be made publicly
available through the CDS service.

\subsection{Comparison to other catalogs}

\begin{figure*}
\plotone{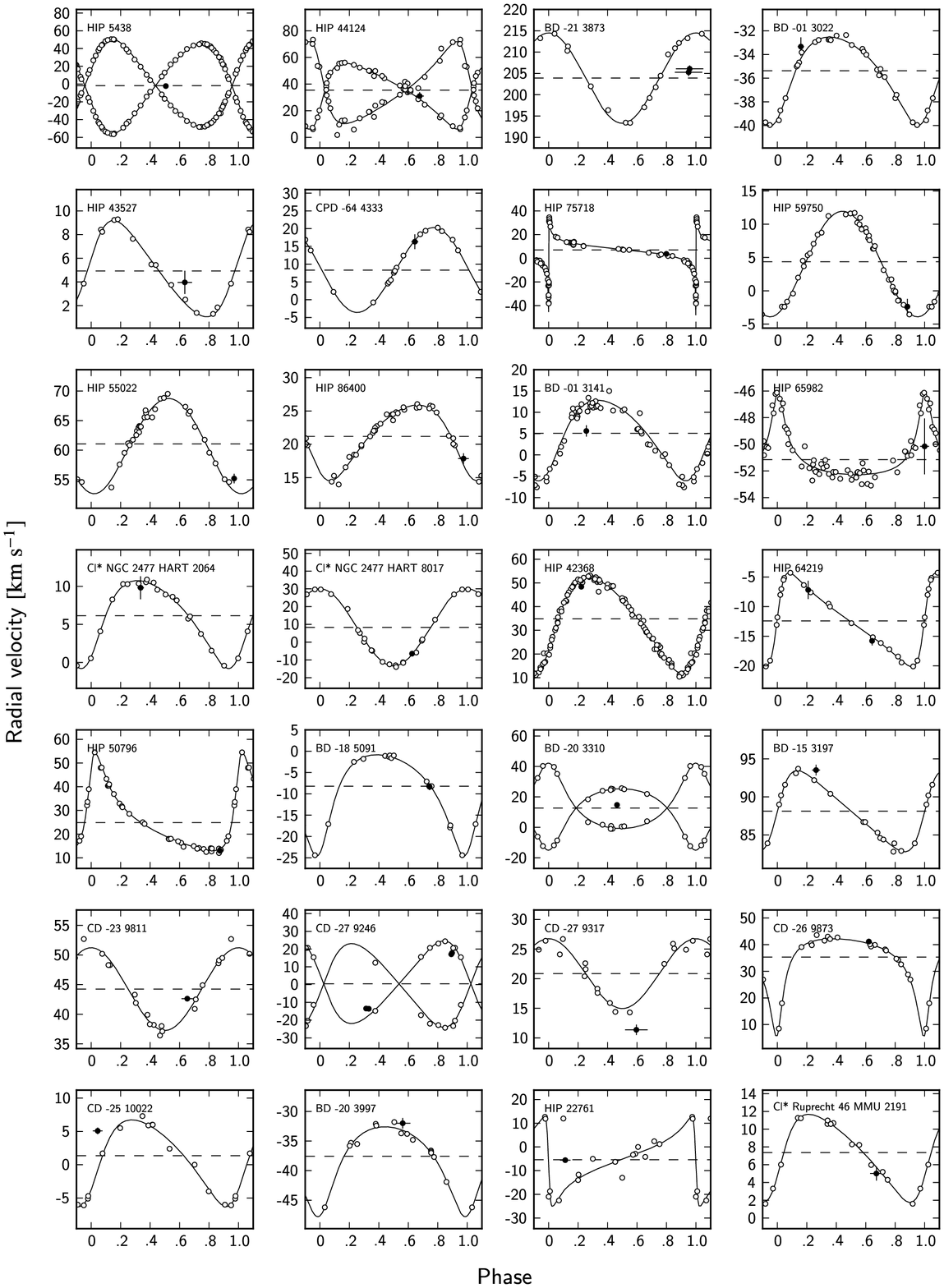}
\figcaption{Spectroscopic orbits from $S_{B^ 9}$ catalog and RAVE RVs (black 
markers) for each star. The phase uncertainty is calculated from a given period 
error. The radial velocity data for HIP~5438 from \citet{1988A&A...196..128A}, 
HIP~44124 from \citet{2002AJ....124.1132G}, BD~-21~3873 from 
\citet{1997A&A...324...97S}, BD~-01~3022, HIP~43527 and CPD~-64~4333 from 
\citet{1998A&AS..131...25U}, HIP~75718 and HIP~86400 from \citet{1991A&AS...91..497T}, 
HIP~59750 and HIP~55022 from \citet{2001AJ....122.3419C}, BD~-01~3141 from \citet{1994Obs...114..231G}, 
HIP~65982 from \citet{2002AJ....124.1144L}, Cl*~NGC~2477~HART~2064 and 
Cl*~NGC~2477~HART~8017 from \citet{2004A&A...423..189E}, HIP~42368 from 
\citet{2004Obs...124...97G}, HIP~64219 from \citet{1991A&A...248..485D},
HIP~50796 from \citet{2006AJ....131.1022T}, BD~-18~5091 from \citet{2007AN....328..527C}, 
BD~-20~3310, BD~-15~3197, CD~-23~9811, CD~-27~9246, 
CD~-27~9317, CD~-26~9873, CD~-25~10022, BD~-20~3997 and HIP~22761 from 
\citet{2006MNRAS.371.1159G} and Cl*~Ruprecht~46~MMU~2191 from 
\citet{2007A&A...473..829M}.
\label{plot_orbits}}
\end{figure*}

We compared the list of identified SB1 candidates to the list of known photometric variables 
in the VSX catalog \citep[Version 02-Jan-2011]{2006SASS...25...47W} and to the 
ninth catalog of spectroscopic binary orbits \citep[$S_{B^9}$;][]{2004A&A...424..727P}. Within 
the VSX there are 36 stars matching our list. Out of those, 10 
are pulsating variables, 14 eclipsing binaries and 20 unclassified variables. 
Their positions on the $\{T_\mathrm{eff},\log g\}$ diagram are shown in Figure~\ref{plot_hr}. RV variations 
measured for pulsating variables are generally lower than $30\,\mathrm{km\ s^{-1}}$
\citep{2010aste.book.....A}, but can be as high as $\sim 100\,\mathrm{km\ s^{-1}}$ for
some types \citep{1998AstL...24..815G,1949CMWCI.757....1S}. In our case, the two leftmost green 
dots in the upper right panel of Figure~\ref{plot_hr} are an RR Lyrae star (hotter) and 
a Cepheid with the RV amplitude of at least 
$21\,\mathrm{km\ s^{-1}}$ and $46\,\mathrm{km\ s^{-1}}$, respectively. The rest 
of the pulsating variables mostly lie above the dwarf-giant border. Of course, 
there is a possibility that some of the rest of the identified SB1 candidates 
are pulsating variables rather than binary stars. Conversely, eclipsing binaries 
and potential SB2 candidates are generally (taking atmospheric parameter
uncertainties into account) among main sequence dwarfs. This supports the 
assumption from Paper I and also agrees with the sample of eclipsing 
binaries from \citet{2010A&ARv..18...67T} where the majority of all analyzed 
systems are pairs of main-sequence stars.

The comparison of the whole RAVE sample to the $S_{B^9}$ catalog yielded 56 
matches but only three of those stars were observed more than once, two of them 
twice, and one four times. For the sake of testing the accuracy of RVs, 
single observations were compared to known orbits as well. We selected 26 
orbital solutions for which the difference between the published periastron time 
and the date of the RAVE observation were closer than 100 periods apart in order 
to avoid errors caused by period propagation. There are two exceptions with a difference of 
343 and 448 periods where the correspondence between the RAVE measurements and
the RV curve was still acceptable. In other cases several
thousand periods have passed between the given periastron time and the time of the RAVE measurement. Orbital 
solutions with RAVE measurements are shown in Figure~\ref{plot_orbits}. There are 
four known SB2 systems among the selected sample. RAVE spectra of HIP~5438 and 
HIP~44124 were recorded close to the half phase and so both RVs are near the 
systemic $\gamma$ velocity. The RV amplitudes of both systems are large enough that both 
systems would be detected as SB2 if observed at more favorable phases. System 
BD~-20~3310 was observed close to the apastron but again the measured RV is near 
the center-of-mass RV because spectra of both components contribute to the 
composite so the peak of the correlation function lies in the middle of both 
components. System CD~-27~9246 was observed four times, but coincidentally at 
only two different phases. All four RVs correspond to one component of the system. Other RAVE RVs 
are close to the predicted values and in most cases within the error bars, 
giving additional reassurance to the error estimates. In three cases, 
BD~-21~3873, BD~-01~3141, and HIP~22761, the measured RVs lie very close to the 
$\gamma$ velocities of the system, very similar to the three cases of known SB2 systems.
The reason for this seems to be different in each case. A poor match in the
case of HIP~22761 (HD 31341) is most likely caused by confusion with another 
star due to an erroneous entry in the catalog. The $S_{B^9}$ catalog cites \citet{2006MNRAS.371.1159G} as the source, 
but the paper actually reports about the observation of the star HD~31341B, also known as HIP~22766.
The mismatch in BD~-21~3873 could be caused by an improper phase determination of the two RAVE measurements,
since the phase inaccuracy in this case is the largest of all examples.
In the case of BD~-01~3141, a possible secondary component might only be observable at
longer wavelengths because of its potentially low surface temperature. The reference paper
\citep{1994Obs...114..231G} states that all observations were performed on instruments
capable of reaching wavelengths up to $\sim5200\,$\r{A} which is relatively far from RAVE's
spectral domain so there is a chance that the additional component was not
present there. 

\section{Discussion and conclusions}

We analyzed 20,027 stars observed multiple times by the RAVE 
spectroscopic survey in the $\sim 6$ year span between 2004 and 2010 with RV
and atmospheric parameter estimates provided by the RAVE parameter 
estimation pipeline. Prior to the RV analysis we classified all 
spectra and filtered out those with peculiar features (e.g., emission line 
objects, double-lined binaries, spectra with observational errors, etc.) whose 
measured parameters are not reliable, obtaining a list of only normal single 
stars. In order to detect stars with variable RVs, we defined a 
variability criterion (Equation~(\ref{eq_erf})), assuming that RV 
errors are Gaussian and taking the RV error and 
zero-point shift of individual measurements into account. In the observed sample, we identified 1333 objects with large 
enough changes in radial velocities to be identified as SB1 candidates. As 
summarized in Table~\ref{table_sb1}, the fraction of discovered SB1 candidates is 
a function of the number of observations. Only 6 \% of twice observed stars were 
detected as SB1 candidates. This fraction grows with the number of 
observations and saturates at around 10\%-15\% for stars with five or six 
observations, depending on the selected lower limit of the variability criterion. 
This number represents the lower limit for the overall 
fraction of stars with variable RVs present in the RAVE sample.

The distribution of maximal differences between the RVs for 
repeatedly observed stars (Figure~\ref{plot_rv_dist}) has a strong peak just 
below $10\,\mathrm{km\, s^{-1}}$. At smaller velocities the number of detected 
SB1 candidates quickly vanishes. The reason is a restriction on the
$p_\mathrm{log}$ criterion that is required to be greater than 3 for
sufficiently confident variability confirmation. Even in cases with the lowest values
of RV errors (Figure~\ref{plot_rv_errors}) this criterion is not met if the two measurements are
closer together than $\sim3.5\,\mathrm{km\,s^{-1}}$ (Figure~\ref{plot_erf_prob}) 
which explains the sharp falloff below $4\,\mathrm{km\, s^{-1}}$. The same distribution
was plotted for stars where $p_\mathrm{log}>4$ and the falloff was even stronger
there which gives additional confirmation to the source of this feature.
It should be noted that zero-point shifts (Figure~\ref{plot_rv_errors}) are in 
most cases well below the velocities where the detection becomes significant and 
are not the source of inaccuracies. On the other side of the distribution,
there is an evident exponential falloff of detected candidates. This feature is 
expected to be real and not a selection effect. Systems with larger RV 
amplitudes most likely have shorter periods and therefore the detection 
efficiency for these cases is proportionally higher than at the lower velocity 
end. The distribution also extends into the SB2 region for which it 
was shown (Paper I) that the detection efficiency is very high at separations 
above $50\,\mathrm{km\, s^{-1}}$. It is hard to infer the periods of the 
SB1 candidates since mostly only two observations are available, but from the 
distribution of the time spans between observations it can be concluded that 
periods are most likely not significantly longer and so fall into the short end 
of the period distribution given in \citet{1991A&A...248..485D} or 
\citet{2010ApJS..190....1R}. A detailed population study of both types of binaries
(SB1 and SB2) and their influence on the general RV distribution of the
RAVE stars will be a subject of the third paper regarding binaries in the 
RAVE survey.

The $\{T_\mathrm{eff},\log g\}$ diagram of SB1 candidates derived from the spectroscopic parameters of 
stars gives a similar picture as RAVE's general population.
The distribution of metallicities of SB1 candidates is the
same as the distribution of metallicities of the general population
which means that binaries are well mixed among the field stars. 
Similarly to the overall RAVE sample, the distribution of effective temperatures 
has two distinct peaks that correspond to main sequence stars and to the red clump giants
with masses only slightly larger than the former group.
There is a deficiency of SB1 candidates from the giant 
group compared to the dwarfs. It can be explained by comparing the typical radii of
stars in both groups. Radii of giants in the observed sample are roughly an order
of magnitude larger than the radii of main sequence stars. Therefore, binary orbits
in which giants can exist (due to Roche lobe limits) must be on average larger 
than the orbits of main sequence stars so the orbital periods are longer and 
consequently RV shift smaller and more unlikely to be detected. 

Comparison of the SB1 candidate list with the VSX catalog of photometrically 
variable stars yielded several hits with some stars being pulsating variables 
rather than spectroscopic binaries. The RV amplitudes of some types of 
pulsating variables can be as high as few $10\,\mathrm{km\, s^{-1}}$. There is some
probability that some of the stars from our SB1 candidate list are pulsating but 
it should be negligibly low. For example, the fraction of Cepheids and RR Lyrae stars
(variables with high RV variations)
in the \textit{Hipparcos} catalog \citep{1997A&A...323L..49P} is less than 0.4 \%. The selection of stars in the RAVE's input catalog 
is unbiased (with the exception of stellar magnitudes) so most of the observed 
stars are not in any unstable transiting phases of their lives.
Among the photometrically variable stars are also several 
eclipsing binaries. Most of them lie close to the main sequence on the HR diagram. 
We also identified some potential SB2 candidates in our SB1 candidate list. 
They cannot be confirmed since they do not have a clear 
double-lined signature in their spectra, but time dependent asymmetries in 
spectral lines indicate that higher resolution spectra might be able to resolve two 
components. All of these objects also lie close to the main sequence which 
supports the argument that SB2 objects mostly consist of two main sequence stars and
pairs of equally bright giants forming an SB2 system are rare.

There are 26 stars in the catalog of spectroscopic binary orbits ($S_{B^9}$) for 
which RAVE has at least one observation and the difference between the given 
time of periastron and the RAVE observation time is shorter than 100 epochs to 
ensure that the propagated period error is small enough. In most cases the 
agreement between the calculated RV curves and RAVE RVs is within the error 
estimates which confirms the precision of the RVs. Among the selection are four
known SB2 objects. In three of those cases, the RAVE RV is close to 
the $\gamma$ velocity of the system which is expected because the peak of the 
correlation function that determines the RV is between both 
Doppler-shifted components. Conversely, in three other cases that are cataloged as SB1 
objects, RAVE radial velocities are close to the systemic velocity and not
near the calculated RV curve. This might implicate the presence of a still 
undiscovered bright secondary component.

\acknowledgements
Funding for RAVE has been provided by: the Australian Astronomical
Observatory; the Leibniz-Institut für Astrophysik Potsdam (AIP); the Australian
National University; the Australian Research Council; the French National Research
Agency; the German Research Foundation; the European Research Council 
(ERC-StG 240271 Galactica); the Istituto Nazionale di
Astrofisica at Padova; The Johns Hopkins University; the National Science
Foundation of the USA (AST-0908326); the W. M. Keck foundation; the
Macquarie University; the Netherlands Research School for Astronomy; the
Natural Sciences and Engineering Research Council of Canada; the Slovenian
Research Agency; the Swiss National Science Foundation; the Science \&
Technology Facilities Council of the UK; Opticon; Strasbourg Observatory;
and the Universities of Groningen, Heidelberg and Sydney. The RAVE web site is at
http://www.rave-survey.org.

\end{document}